\RequirePackage{rotating}

\documentclass[12pt]{article}

\usepackage[margin=1in]{geometry}
\usepackage{amsmath,amssymb,amsfonts}
\usepackage{hyperref}

\usepackage{times}
\usepackage[T1]{fontenc}
\usepackage{graphics, graphicx}
\usepackage{color}
\usepackage[FIGTOPCAP]{subfigure}
\usepackage{textpos}
\usepackage{natbib,float}
\usepackage{authblk}
\usepackage[english]{babel}
\usepackage{caption}

\DeclareGraphicsExtensions{.pdf,.png,.jpg}

\usepackage{setspace}
\allowdisplaybreaks

\def\BibTeX{{\rm B\kern-.05em{\sc i\kern-.025em b}\kern-.08em
		T\kern-.1667em\lower.7ex\hbox{E}\kern-.125emX}}

\title{A latent spatial factor approach for synthesizing opioid associated deaths and treatment admissions in Ohio counties}

\date{}

\author[1]{Staci A. Hepler}
\author[2]{Erin McKnight}
\author[3]{Andrea Bonny}
\author[4]{David Kline}
\affil[1]{Department of Mathematics and Statistics, Wake Forest University}
\affil[2,3]{Nationwide Children's Hospital}
\affil[4]{Center for Biostatistics, The Ohio State University}

\begin{document}

\maketitle

\begin{abstract}
	\textbf{Background:} Opioid misuse is a major public health issue in the United States and in particular Ohio.  However, the burden of the epidemic is challenging to quantify as public health surveillance measures capture different aspects of the problem.  Here we synthesize county-level death and treatment counts to compare the relative burden across counties and assess associations with social environmental covariates.
	\textbf{Methods:} We construct a generalized spatial factor model to jointly model death and treatment rates for each county.  For each outcome, we specify a spatial rates parameterization for a Poisson regression model with spatially varying factor loadings. We use a conditional autoregressive model to account for spatial dependence within a Bayesian framework. 
	\textbf{Results:} The estimated spatial factor was highest in the southern and southwestern counties of the state, representing a higher burden of the opioid epidemic.  We found that relatively high rates of treatment contributed to the factor in the southern part of the state; whereas, relatively higher rates of death contributed in the southwest.  The estimated factor was also positively associated with the proportion of residents aged 18-64 on disability and negatively associated with the proportion of residents reporting white race. 
	\textbf{Conclusions:} We synthesized the information in the opioid associated death and treatment counts through a spatial factor model to estimate a latent factor representing the consensus between the two surveillance measures.  We believe this framework provides a coherent approach to describe the epidemic while leveraging information from multiple surveillance measures. \newline 
\end{abstract}

\section*{Background}
Opioid misuse is currently a major public health issue in the United States because of its high prevalence and associated morbidity and mortality \citep{WH2011}. When asked about use in the past month, approximately 4 million persons reported nonmedical use of a prescription opioid and 400,000 reported heroin use \citep{Brady2016,Dart2015,CBHSQ}.  From 2000 to 2014, the rate of opioid overdose deaths has increased 200\% \citep{Rudd2016a} and are now the leading cause of injury-related death in the United States \citep{Chen2015}.

In Ohio, the toll of the epidemic has been particularly severe. In 2016, Ohio ranked second for highest overdose death rate \citep{Hedegaard2017}. The death rates have only continued to increase in Ohio as fentanyl has penetrated the state \citep{Daniulaityte2017}.  This has led to various policy initiatives aimed at directing resources for treatment to affected areas, particularly southern Ohio \citep{Governor2012}.

When studying the epidemiology of the opioid epidemic, one particularly challenging aspect is choosing how to quantify its local burden.  Direct evidence on opioid misuse at the county level is often difficult to obtain because large ongoing public health surveys are not often designed to estimate rates at the county level and illicit drug use is likely to be underreported \citep{Palamar2016}.  Surveys using complex designs, like respondent driven sampling \citep{Handcock2014}, have been designed to address some of these questions but are often quite resource and time intensive \citep{Platt2006}.  

Rather than use survey data, we elect to take advantage of surveillance data that are routinely collected by the state of Ohio.  The state monitors opioid associated deaths and treatment admissions at the county level.  Each outcome provides related but slightly different information regarding opioid misuse in a county. Rather than choose a single marker as a proxy for the burden of the opioid epidemic, we jointly modeled both outcomes in an attempt to gain a more comprehensive understanding of the epidemic.  By doing so, we can leverage associations between outcomes to improve estimation and make joint inferences \citep{Martinez-Beneito2017}.  More interestingly, we can extract common features from both counts to construct a spatial latent factor \citep{Wall2003,Neeley2014}.  Like other confirmatory factor models, we can assign the factor an interpretation based on its indicators. In this case, the indicators are opioid associated treatment and death counts so we choose to interpret the latent factor as the unobserved ``burden'' of the opioid epidemic for each county.  This latent factor can then be used to evaluate the relative burden across the state and assist policy makers and public health professionals in targeting counties that are most in need of intervention.

The computational objective of this study is to utilize the common conditional autoregressive (CAR) framework \citep{Besag1974}, which allows for handling of spatial dependence, and a spatial factor model \citep{Wall2003} to better characterize the opioid epidemic in individual counties in the state of Ohio. The epidemiological objective is to use this modeling approach to synthesize two available surveillance measures to assess the county-level burden of opioid misuse.  We also examine ecological associations of the level of burden with sociodemographic characteristics.

\section*{Methods}

\subsection*{Data}
Our model is based on routinely collected surveillance data from the state of Ohio.  The state monitors opioid associated deaths and treatment admissions for all of Ohio's 88 counties. For this analysis, we used aggregate counts from the three most recent available years, 2013-2015.  Death counts are publicly available from the Ohio Department of Health website (http://publicapps.odh. ohio.gov/EDW/DataCatalog) and were obtained from the Ohio Public Health Data Warehouse Ohio Resident Mortality Data. We included all resident deaths where poisoning from any opiate is mentioned on the death certificate.  Deaths are counted in the county where the decedent resided at the time of death regardless of where the death occurred.  ICD-10 multiple cause codes T40.0-T40.4 and T40.6 present on the death certificate denote poisoning from any opiate. Raw observed death rates for each county are displayed in Supplement Figure 1(A).

We obtained treatment admission counts by patient county of residence through a data use agreement with the Ohio Department of Mental Health and Drug Addiction Services. Treatment admissions were identified through the diagnostic codes shown in Supplement Table 1 and include any residential, intensive outpatient, or outpatient treatment for opioid misuse.  Patients who report to hospitals or any other medical facility to receive treatment for overdose or other complications from opioid misuse are not included in this count.  Patients with multiple admissions are only counted once.  Data were provided separately for those under and over age 21, but we will only consider the total counts in this analysis.  Counts under 10 were suppressed or censored as a matter of state policy, which impacts four counties in this data set.  Raw observed treatment rates are shown in Supplement Figure 1(B) with censored counties marked in bold. We also note a moderate, positive correlation between the death and treatment rates within fully observed counties ($r=0.46$). A scatter plot of rates is shown in Supplement Figure 1(C).

We note that Van Wert County was excluded from the analysis due to a data quality issue.  Thus, 87 counties were used in the analyses that follow.

We obtained county level covariate information to examine associations between social environmental factors and the latent burden of the opioid epidemic. We used the 2015 5-year estimates from the United States Census Bureau's American Community Survey to provide county population and demographic characteristics.

\subsection*{Statistical Considerations}
For this study, we have bivariate count observations of opioid associated deaths and treatment admissions for 87 of Ohio's counties.  We elect to take a Bayesian approach for analyzing county level areal data. We will briefly address the statistical approach here and will defer full details of the model to the Supplement.

Our primary goal of the analysis is to synthesize the information from the rates of death and treatment in each county through a generalized common spatial factor model \citep{Wall2003}. Thus, we assume that there is an underlying spatial factor or latent variable that drives both death and treatment rates and also accounts for spatial dependence with factors of neighboring counties.  The spatial factor is shared across both outcome models within a county as in a structural equation model \citep{Muthen2012} or shared latent variable model \citep{Dunson2003}.  This assumes that there are common underlying conditions in a county that are associated with both death and treatment rates.  In this case, we interpret those conditions as the "burden" of the opioid epidemic in a particular county.  We can then look for associations of the burden with social environmental covariates by including them as fixed effects in the mean structure of the spatial factor.

The joint model for death and treatment rates is specified as a generalized common spatial factor model \citep{Wall2003} for Poisson outcomes with spatially varying loadings \citep{Neeley2014}.  For each Poisson model, we use the spatial rates parameterization \citep{Cressie2005,Cressie2011}. The spatial structure in the latent factor and the loadings are characterized using an intrinsic CAR model \citep{Banerjee2004,Besag1974}. Uncorrelated heterogeneity is modeled using independent latent factors for each outcome.  We account for the censored treatment counts through an adaptation of a censored generalized Poisson regression model \citep{Famoye2004} for the case of interval censoring.

The full technical model specification is provided in the Supplement.  We also address the specification of prior distributions, identifiability of the model, and computational details in the Supplement.  The model was fit using a Markov chain Monte Carlo algorithm that was implemented in MATLAB version 2015a and run using a single core of a 128GB node on a high performance computing cluster.

\section*{Results}

\begin{figure}
	\centering
	\subfigure[Log Standardized Mortality Ratio]{
		\label{fig:est_death}
		\includegraphics{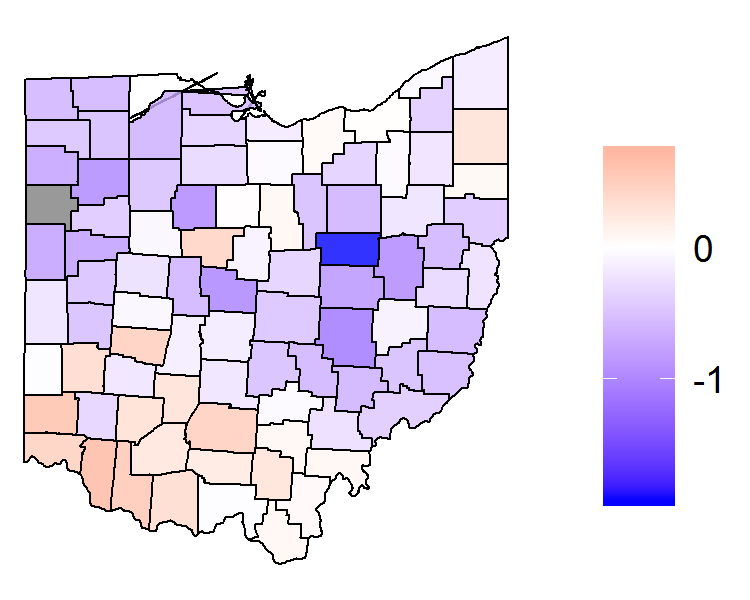}
	}
	\subfigure[Log Standardized Treatment Rate Ratio]{
		\label{fig:est_trt}
		\includegraphics{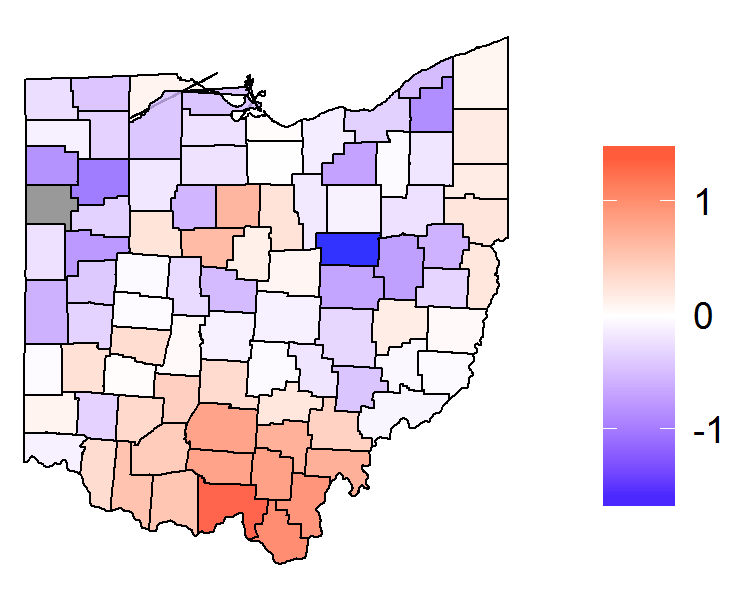}
	}
	\subfigure[Estimated Spatial Factors]{
		\label{fig:est_v}
		\includegraphics{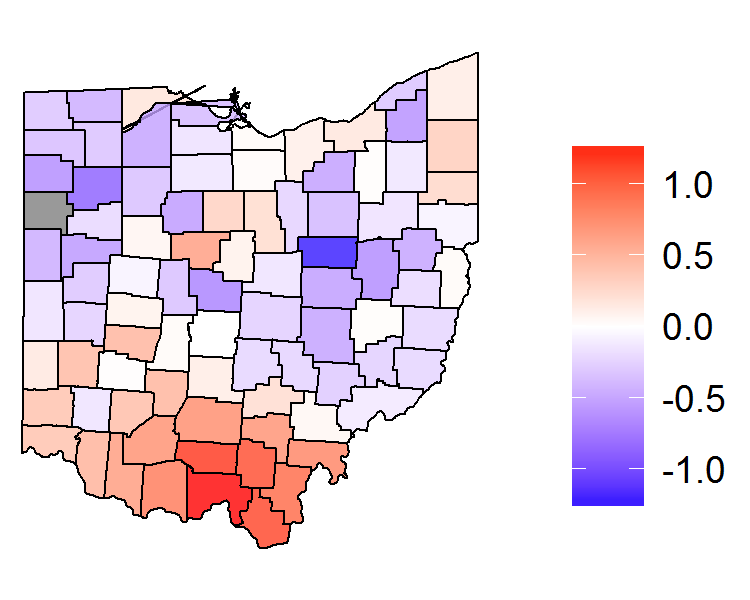}
	}
	\subfigure[Estimated Loadings]{
		\label{fig:est_alpha}
		\includegraphics{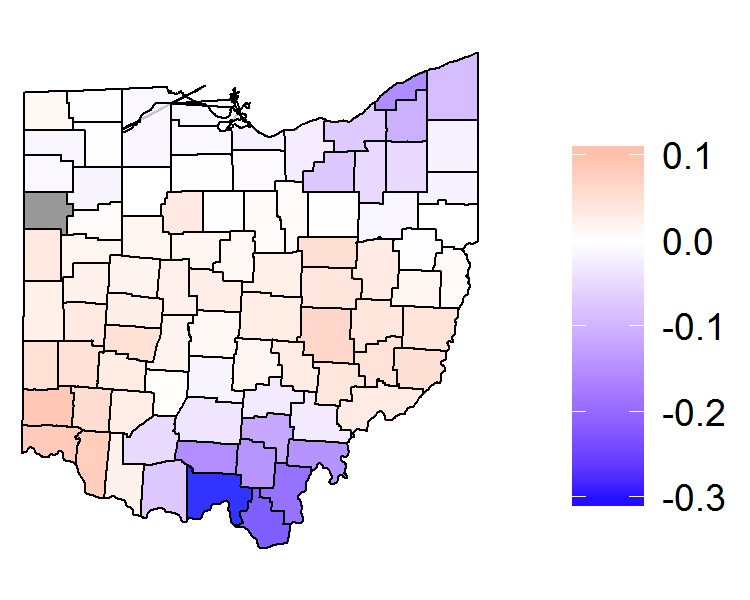}
	}
	\caption[Model Estimates]{Posterior mean estimates of model parameters for each county.}
	\label{fig:est}
\end{figure}

The posterior mean estimates of the log standardized mortality ratio (SMR) are shown in Figure \ref{fig:est}\subref{fig:est_death}.  Each ratio reflects a comparison to the overall state opioid overdose death rate, which is 57 deaths per 100,000 residents over the three year period under study. We observe the highest rates of opioid overdose death in the southwestern portion of the state near Cincinnati.  We see approximately average rates of death in southern Ohio and in the northeastern part of the state near Cleveland. Below average rates are observed in the northwestern and eastern portions of the state. 

Figure \ref{fig:est}\subref{fig:est_trt} shows the log standardized treatment rate ratio for each county. Each ratio is compared to the overall state opioid misuse treatment admission rate of 563 admissions per 100,000 residents over this three year period.  We see the highest rates of opioid misuse treatment admissions in southern Ohio which is consistent with an initial state effort to direct resources to that area \citep{Governor2012}.  We also see slightly elevated rates along the states eastern border and again see below average rates in the northwestern part of the state.

A map of the posterior mean estimates of the spatial factor for each county is shown in Figure \ref{fig:est}\subref{fig:est_v}. The spatial factor illustrates the consensus between the death and treatment rates for each county. This synthesis is evident when comparing Figure \ref{fig:est}\subref{fig:est_v} to Figures \ref{fig:est}\subref{fig:est_death} and \ref{fig:est}\subref{fig:est_trt}. The spatial factor has been constructed such that high values reflect a higher burden of the opioid epidemic. We note that the scale of the latent factor is arbitrary so that only relative comparisons are meaningful. When looking at Figure \ref{fig:est}\subref{fig:est_v}, we see that the areas of highest burden are in the southern and southwestern portions of the state. We see slightly elevated burdens in the northeastern part of the state.  We note below average burdens in the northwest part of the state as well as in the east central portion.

The posterior mean estimates of the  scaled loadings  are shown in Figure \ref{fig:est}\subref{fig:est_alpha}. The technical purpose of the loadings is to allow the covariance between the death and treatment rates to vary across space.  However, we can also interpret a scaled version of the loadings to gain additional insights \cite{Neeley2014}. For loadings that are close to zero, this indicates that death and treatment rates are contributing roughly the same to the factor. For loadings greater than zero, death rates exert more influence over the factor, and for loadings less than zero, treatment rates are more influential. More simply, the loadings highlight areas with treatment or death rates that are more extreme than we would otherwise expect.  We see loadings greater than zeros in southwestern and eastern Ohio where rates of death are high compared to the treatment rates. In contrast, in southern Ohio, we see negative loading estimates which represent much higher rates of treatment than expected based on death rates.

Table \ref{table:trt_effects} displays the posterior mean estimated regression coefficients and 95\% credible intervals for standardized covariates included in the mean of the spatial latent factor.  All estimated associations are conditional on the other covariates in the model.  We observe credible intervals that do not cross zero for the proportion of white residents and the proportion of residents aged 18-64 on disability.  Higher proportions of residents on disability are associated with higher values of the latent factor and thus higher burdens of opioids.  Higher proportions of white residents are associated with lower burdens of opioids.  A map of the proportion of white residents is shown in eFigure 2.  We do not observe evidence of associations between burden of opioids and median age or log of median household income. 

\begin{table}
	\caption{\label{table:trt_effects}Posterior mean estimated regression coefficients and 95\% credible intervals capturing the association between each standardized covariate and the estimated latent factor.}
	\begin{tabular}{lrr}
		\hline
		Variable & Estimated Coefficient & 95\% Credible Interval\\
		\hline
		Median Age & -0.014 & (-0.103, 0.074) \\
		Proportion White & -0.082 & (-0.155, -0.008) \\
		Log of Median Household Income & -0.036 & (-0.151, 0.075) \\
		Proportion Aged 18-64 on Disability & 0.322 & (0.158, 0.483) \\
		\hline
	\end{tabular}
\end{table} 

\section*{Conclusion}
In this paper, we jointly modeled opioid associated deaths and treatment admissions and synthesized their information through a latent spatial factor.  We interpreted the latent spatial factor as the burden of opioids in each county and examined associations between the burden and social environmental covariates.  We observed that the county burden is positively associated with the proportion of county residents aged 18-64 on disability and negatively associated with the proportion of the county that reports their race as white. We were also able to identify counties in the state with higher or lower than average levels of relative burden.

One major advantage of our approach is the ability to synthesize information from multiple outcomes related to opioid misuse to arrive at more comprehensive estimates than if we looked at either outcome in isolation. Since opioid misuse is an illicit activity, it is difficult to measure directly within a surveillance context so it is common to examine proxies like overdose death and treatment admissions.  However, there can be a desire to extend interpretations beyond the specific proxy to infer about a more general assessment of the severity of the epidemic or relative rates of misuse.  This implicitly involves assuming a direct correlation between the proxy and misuse that is constant over space.  However, we observe different spatial patterns for death rates and treatment rates in our analysis which means that the implicit assumption cannot be simultaneously true for both rates.  Instead, we believe that the truth is likely to lie in between and by leveraging the information contained in multiple outcomes, we can obtain an estimate that better reflects heterogeneity across space and extracts the commonalities across outcomes. Thus, rather than making conclusions specific to an outcome or making strong assumptions regarding the relationship between the proxy and misuse, we instead coherently incorporate information from both outcomes in our estimate of the burden of the epidemic.

We ultimately believe that our estimates of burden provide a more relevant marker for policy makers of the relative severity of the epidemic across counties in the state.  By synthesizing multiple proxy outcomes, we are less likely to be misled by features that are specific to any one outcome.  Instead, if we have chosen reasonable proxy outcomes, the consensus between them should provide a better marker of the underlying driver of the outcomes or the burden of the epidemic.  As policy makers allocate resources, areas with the highest burden should be prioritized for additional resources to stem the tide of the epidemic. The results of this analysis suggest that new treatment resources should be allocated to counties where overdose is contributing more to the burden as in southwestern Ohio.  Significant resources have already been allocated to southern and southeastern Ohio \citep{Governor2012}, which is evident by the relatively high rates of treatment.  Our analysis suggests that such efforts should expand by focusing next on southwestern Ohio.  Resources could include tools to reduce overdose deaths like intranasal naloxone or efforts to increase rates of treatment through improved access and novel interventions.

Our approach here serves as the foundation for future modeling and a proof of concept when applied to the opioid epidemic.  We can extend our model to incorporate additional surveillance data as it becomes available. Likewise, we can also extend this framework into the spatio-temporal setting. Additional sources of data and the inclusion of time will introduce additional statistical challenges as there will be other sources of dependence that require modeling and additional assumptions will likely be required about the joint set of surveillance outcomes.   Our approach to this analysis is also not limited to this particular application.  We advocate for the use of joint models and spatial factor models in other similar situations where multiple proxies may be readily available but the true underlying burden of the disease or disorder may be difficult or impossible to obtain.  This approach provides a principled model for synthesizing the information in the multiple outcomes into a readily interpretable relative comparison across space.

There are a few overarching limitations to our analysis.  We utilized state surveillance data on deaths and treatment admissions associated with opioids.  Death count data is derived from reporting on death certificates which can be reported incorrectly \citep{Slavova2015}.  We did not account for this in our analysis. We also only had aggregate, areal data for this analysis and thus conclusions are limited to the county level and cannot be extended to the person level due to the ecological fallacy \citep{Piantadosi1988}.  In addition, our data were aggregated across time so we are not able to describe any temporal trends. 

In conclusion, we have demonstrated an approach using spatial factor models to synthesize multiple outcomes associated with the opioid epidemic to estimate the latent burden of opioids across Ohio counties. We believe this analysis provides a valuable tool for policy makers as they allocate resources to continue to combat the opioid epidemic.  This framework provides a coherent, model-based approach for putting several of the pieces of the puzzle together to gain a more complete picture of the spatial epidemiology of opioid misuse in Ohio.

\bibliography{r21_refs} 

\begin{thebibliography}{10}

\bibitem{WH2011}
{Office of National Drug Control Policy Executive, Office of the President of
  the United States} . Epidemic: responding to {A}merica's prescription drug
  abuse crisis  Internet 2011.
\newblock
  {http://www.whitehouse.gov/sites/default/files/ondcp/policy-and-research/rx\_abuse\_plan.pdf}.

\bibitem{Brady2016}
Brady K.~T., McCauley J.~L., Back S.~E.. Prescription Opioid Misuse, Abuse, and
  Treatment in the {U}nited {S}tates: An Update.  {\it The American journal of
  psychiatry. } 2016;173:18--26.

\bibitem{Dart2015}
Dart R.~C., Surratt H.~L., Cicero T.~J., et al. Trends in opioid analgesic
  abuse and mortality in the {U}nited {S}tates.  {\it The New England journal
  of medicine. } 2015;372:241--8.

\bibitem{CBHSQ}
{Center for Behavioral Health Statistics and Quality} . Behavioral health
  trends in the {U}nited {S}tates: {r}esults from the 2014 National Survey on
  Drug Use and Health  Internet 2015.
\newblock {http://www .samhsa .gov/ data/ sites/ default/ files/
  NSDUH-FRR1-2014/ NSDUH-FRR1-2014 .pdf}.

\bibitem{Rudd2016a}
Rudd R.~A., Aleshire N., Zibbell J.~E., Gladden R.~M.. Increases in Drug and
  Opioid Overdose Deaths--{U}nited {S}tates, 2000-2014.  {\it MMWR. Morbidity
  and mortality weekly report. } 2016;64:50--51.

\bibitem{Chen2015}
Chen L.H., Hedegaard H., Warner M.. QuickStat: Rates of death from drug
  poisoning involving opioid analgesics-{U}nited {S}tates, 1999-2013  {\it
  MMWR. } 2015;64:32.

\bibitem{Hedegaard2017}
Hedegaard H., Warner M., Minino A.M.. Drug overdose deaths in the United
  States, 1999-2016  {\it NCHS Data Brief. } 2017;294.

\bibitem{Daniulaityte2017}
Daniulaityte R., Juhascik M.~P., Strayer K.~E., et al. Overdose Deaths Related
  to Fentanyl and Its Analogs - {O}hio, {J}anuary-{F}ebruary 2017.  {\it MMWR.
  Morbidity and mortality weekly report. } 2017;66:904--908.

\bibitem{Governor2012}
{Governor's Cabinet Opiate Action Team} . Attacking {O}hio's opiate epidemic
  Online; accessed 6-September-2017 2012.
\newblock {http://mha.ohio.gov/}.

\bibitem{Palamar2016}
Palamar Joseph~J., Shearston Jenni~A., Cleland Charles~M.. Discordant reporting
  of nonmedical opioid use in a nationally representative sample of {US} high
  school seniors  {\it The American Journal of Drug and Alcohol Abuse. }
  2016;42:530--538.

\bibitem{Handcock2014}
Handcock Mark~S., Gile Krista~J., Mar Corinne~M.. Estimating hidden population
  size using Respondent-Driven Sampling data  {\it Electronic Journal of
  Statistics. } 2014;8:1491--1521.

\bibitem{Platt2006}
Platt Lucy, Wall Martin, Rhodes Tim, et al. Methods to Recruit Hard-to-Reach
  Groups: Comparing Two Chain Referral Sampling Methods of Recruiting Injecting
  Drug Users Across Nine Studies in {R}ussia and {E}stonia  {\it Journal of
  Urban Health. } 2006;83:39--53.

\bibitem{Martinez-Beneito2017}
Martinez-Beneito M.~A., Botella-Rocamora P., Banerjee S.. Towards a
  multidimensional approach to {B}ayesian disease mapping  {\it Bayesian
  Analysis. } 2017;12:239--259.

\bibitem{Wall2003}
Wall M.~M., Wang F.. Generalized common spatial factor model  {\it
  Biostatistics. } 2003;4:569--582.

\bibitem{Neeley2014}
Neeley E.S., Christensen W.F., Sain S.R.. A Bayesian spatial factor analysis
  approach for combining climate model ensembles  {\it Environmetrics. }
  2014;25:483--497.

\bibitem{Besag1974}
Besag Julian. Spatial Interaction and the Statistical Analysis of Lattice
  Systems  {\it Journal of the Royal Statistical Society. Series B
  (Methodological). } 1974;36:192--236.

\bibitem{Muthen2012}
Muthen B., Asparouhov T.. Bayesian structural equation modeling: a more
  flexible representation of substantive theory.  {\it Psychological methods. }
  2012;17:313--35.

\bibitem{Dunson2003}
Dunson David~B.. Dynamic Latent Trait Models for Multidimensional Longitudinal
  Data  {\it Journal of the American Statistical Association. }
  2003;98:555--563.

\bibitem{Cressie2005}
Cressie Noel, Perrin Olivier, Thomas-Agnan Christine. Likelihood-based
  estimation for {G}aussian MRFs  {\it Statistical Methodology. } 2005;2:1--16.

\bibitem{Cressie2011}
Cressie Noel A.~C., Wikle Christopher~K.. {\it Statistics for spatio-temporal
  data}.
\newblock Hoboken, N.J.: Wiley 2011.

\bibitem{Banerjee2004}
Banerjee Sudipto., Carlin Bradley~P., Gelfand Alan~E.. {\it Hierarchical
  modeling and analysis for spatial data}.
\newblock Boca Raton, Fla.: Chapman \& Hall/CRC 2004.

\bibitem{Famoye2004}
Famoye Felix, Wang Weiren. Censored generalized Poisson regression model  {\it
  Computational Statistics and Data Analysis. } 2004;46:547--560.

\bibitem{Slavova2015}
Slavova S., O'Brien D.~B., Creppage K., et al. Drug Overdose Deaths: Let's Get
  Specific.  {\it Public health reports. } 2015;130.

\bibitem{Piantadosi1988}
Pianntadosi Steven, Byar David~P., Green Sylvan~B.. The Ecological Fallacy
  {\it American Journal of Epidemiology. } 1988;127:893-904.

\bibitem{Gelman2006}
Gelman Andrew. Prior distributions for variance parameters in hierarchical
  models (comment on article by {B}rowne and {D}raper)  {\it Bayesian Analysis.
  } 2006;1:515--534.

\bibitem{Roberts2009}
Roberts G.O., Rosenthal J.S.. Examples of adaptive {MCMC}  {\it Journal of
  Computational and Graphical Statistics. } 2009;18:349--367.

\end{thebibliography}
\bibliographystyle{ama}

\section*{Supplementary Material}

\subsection*{Statistical Model}

The generalized common spatial factor model for bivariate Poisson outcomes forms the basis of our model. We use a spatial rates parameterization \citep{Cressie2005,Cressie2011} in order to estimate the relative risk of death and treatment admission for each county compared to the statewide average.  For each county $i=1,...,87$, let $Y_i^D$ be the count of deaths and $Y_i^T$ be the count of treatment admissions.  If county $i$ does not have the treatment admission censored, then we have
\begin{align} \label{eq:model}
Y_i^D|\lambda_i^D \stackrel{ind}{\sim} \text{Poisson}(E_i^D \lambda_i^D)\\
Y_i^T|\lambda_i^T \stackrel{ind}{\sim} \text{Poisson}(E_i^T \lambda_i^T).
\end{align}
Let $P_i$ represent the population in county $i$.  The baseline expected number of deaths in county $i$ is the fixed quantity $E_i^D=P_i\mu^D$ where $\mu^D=\sum_j Y_j^D / \sum_j P_j$ is the statewide average death rate \citep[Section 4.2.6]{Cressie2011}. Likewise, the baseline expected number of treatment admissions for county $i$ is $E_i^T=P_i\mu^T$ where $\mu^T$ is the statewide average treatment rate. Under this parameterization, $\lambda_i^D$ represents the relative risk of death in county $i$ compared to the statewide average and $\lambda_i^T$ represents the relative risk of treatment admission in county $i$ compared to the statewide average. Note that we assume all dependence is modeled through $\lambda_i^D$ and $\lambda_i^T$.

Treatment admission counts that were less than 10 for either adolescents or adults were censored. There are four counties with adolescent counts less than 10 but adult counts greater than 10. For these four counties, we know the true count lies somewhere between the adult count and the adult count plus 9. Thus, we have a situation analogous to interval censoring and can incorporate that information into the likelihood by adapting the censored generalized Poisson regression model \citep{Famoye2004} to the more general case of interval censoring.  Let $c_i$ be an indicator function such that
\begin{align}
c_i=\begin{cases}
0 & \text{total count observed}\\
1 & \text{adolescent count censored}.
\end{cases}
\end{align}
Thus, our data model for both counts of death and treatment admissions leads to the likelihood function
\begin{equation}
\begin{aligned} \label{eq:lik}
L(&\boldsymbol{\lambda}^T, \boldsymbol{\lambda}^D | \textbf{Y}^T, \textbf{Y}^D) = \left[ \prod_{i=1}^{87} f(Y_i^D | \lambda_i^D) \right] \times \\
& \left[ \prod_{i=1}^{87} [f(Y_i^T|\lambda_i^T)]^{I(c_i=0)}[F(Y_i^T+9|\lambda_i^T)-F(Y_i^T-1|\lambda_i^T)]^{I(c_i=1)} \right]
\end{aligned}
\end{equation}
where $f(\cdot|\lambda_i^T)$ and $F(\cdot|\lambda_i^T)$ are the probability mass function and cumulative distribution function, respectively, of a Poisson random variable with mean $E_i^T \lambda_i^T$, $f(\cdot|\lambda_i^D)$ is the probability mass function of a  Poisson random variable with mean $E_i^D \lambda_i^D$, and $I(\cdot)$ is an indicator function. 

Using the canonical log link function for the relative risk, we assume the following joint generalized linear mixed effects model for county $i$:
\begin{equation}
\begin{aligned} \label{eq:loglink}
\text{log}(\lambda_i^D)&=\beta_0^D+\alpha_i^D \nu_i+\epsilon^D_i\\
\text{log}(\lambda_i^T)&=\beta_0^T+\alpha_i^T \nu_i+\epsilon^T_i,
\end{aligned}
\end{equation}
where $\beta_0^D$ and $\beta_0^T$ are intercepts, $\nu_i$ is the factor for county $i$, $\alpha_i^D$ and $\alpha_i^T$ are the factor loadings for death and treatment, respectively, at county $i$, and $\epsilon_i^D$ and $\epsilon_i^T$ are error terms  assumed to be independent, normally distributed with mean zero and variances $\sigma^2_D$ and $\sigma^2_T$, respectively. 

Two of the primary goals of this analysis are estimation of the common spatial factor $\nu_i$ and estimation of covariate effects related to the latent joint ``burden'' represented by $\nu_i$.  As it is shared across the models for death and treatment, it induces correlation between the death and treatment counts within a county.  By adding spatial structure to the vector $\boldsymbol{\nu} = (\nu_1,...,\nu_{88})'$, it also induces spatial dependence on the models for both death and treatment.  We specify an intrinsic conditional autoregressive (CAR) model for $\boldsymbol{\nu}$. That is, we assume the conditional distributions of the latent factor for each county are given by
\begin{equation}
\label{eq:icarcondl}
\nu_i | \nu_{-i} \sim N\left( \bf{X}_i \boldsymbol{\beta} + \frac{1}{w_{i+}} \sum_{j\sim i} \left( \nu_j - \bf{X}_j \boldsymbol{\beta} \right), \frac{\tau^2}{w_{i+}} \right),
\end{equation}
where $\bf{X}_i$ is the $i$th row of a design matrix with no intercept and whose covariates have all been standardized to allow $\beta_0^D$ and $\beta_0^T$ to be interpreted as an overall average. In \eqref{eq:icarcondl}, $w_{i+}$ is the number of neighbors of county $i$ and the summation is over all the neighboring counties. Equation \ref{eq:icarcondl} induces the joint distribution 
\begin{align} \label{eq:car}
\boldsymbol{\nu} \sim N(\bf{X}\boldsymbol{\beta},\tau^2 {\bf{Q}}^{-1}),
\end{align}
where $\bf{Q}$ is the precision matrix given by $\bf{Q} = \bf{D}-\bf{W}$. The matrix $\bf{W}$ is the adjacency matrix whose $(i,j)$th element is one if counties $i$ and $j$ are neighbors and zero otherwise, and $\bf{D} = \text{diag}(w_{i+})$ is the diagonal matrix whose $(i,i)$th element is the number of neighbors of county $i$. The joint distribution \eqref{eq:car} induced by the intrinsic CAR model is not a valid distribution since $\bf{Q}$ is not of full rank; however, it can be used as a prior distribution of a spatial random effect provided a centering constraint such as $1/n \sum_i \nu_i =0$ is imposed \citep{Banerjee2004}. This centering constraint is also necessary to ensure the latent factor is identifiable with the intercepts $\beta_0^D$ and $\beta_0^T$ \citep{Wall2003}. Note that centering the latent factor to be mean zero is reasonable since the covariates in $\bf{X}$ have all been centered. 

As discussed in \cite{Neeley2014}, for the model in equation \eqref{eq:loglink} to be identifiable, we must assume one of the outcomes is used as a reference and all of the loadings for that outcome are constant. Thus, we assume $\alpha_i^T=1$ for $i=1,...,n$. The loadings for death vary spatially and are assumed to follow an intrinsic CAR model with mean one such that if  $\boldsymbol{\alpha}^D = (\alpha_1^D,...,\alpha_n^D)'$, then $$\boldsymbol{\alpha}^D \sim N\left(\boldsymbol{1},\tau^2_D {\bf{Q}}^{-1}\right).$$ 
A centering constraint $1/n \sum_i \alpha_i^D =1$ in enforced so that the intrinsic CAR model can be used as a prior model for the random effect. Centering at one implies loadings that are equal to one correspond to locations where death has a similar influence on the latent factor as treatment. To ease in interpretation, the results presented in the manuscript are the loadings for death at each location divided by the sum of the two outcome loadings at that location. Then, 0.5 is subtracted from these rescaled loadings. After rescaling and centering in this way, loadings of approximately zero indicate locations where the two outcomes have similar influence on the latent factor. Positive values for death indicate locations where death has greater influence on the latent factor than treatment, and negative values indicate locations where treatment has greater influence on the latent factor than death. 

We specify prior distributions for all parameters in the model. The intercepts $\beta_0^D$ and $\beta_0^T$ are assigned independent, non-informative prior distributions that are uniform on the real line. The regression coefficients in the latent factor, $\boldsymbol{\beta}$ are assumed to follow independent normal prior distributions with mean zero and variance 4. 
The variance parameters $\tau^2$, $\tau_2^D$, $\sigma^2_D$, $\sigma^2_T$ are independently uniform on the standard deviation \citep{Gelman2006}. That is, 
\begin{equation}
\begin{aligned} \label{eq:varprior}
\tau &\sim U(0,\infty)\\
\tau_D &\sim U(0,\infty)\\
\sigma_D &\sim U(0,\infty)\\
\sigma_T &\sim U(0,\infty).
\end{aligned}
\end{equation}
The posterior distribution of all unobservable quantities is given by
\begin{equation} 
\begin{aligned} 
\pi(\boldsymbol{\nu}, \boldsymbol{\alpha}^D, \boldsymbol{\epsilon}^D, &\boldsymbol{\epsilon}^T, \beta_0^D, \beta_0^T, \boldsymbol{\beta}^D, \boldsymbol{\beta}^T \tau^2, \tau^2_D \sigma_D^2, \sigma_T^2 \, | \, \bf{Y}^D, \bf{Y}^T) \\
&\propto L(\boldsymbol{\lambda}^T, \boldsymbol{\lambda}^D | \textbf{Y}^T, \textbf{Y}^D) f(\boldsymbol{\nu} | \tau^2) f(\boldsymbol{\alpha}^D | \tau^2_D) f(\boldsymbol{\epsilon}^D | \sigma^2_D) f(\boldsymbol{\epsilon}^T | \sigma^2_T) \\ &\times \pi(\beta_0^D) \pi(\beta_0^T)\pi(\boldsymbol{\beta})   \pi(\tau^2) \pi(\tau^2_D) \pi(\sigma^2_T) \pi(\sigma^2_D),
\end{aligned}
\end{equation}
where $L(\cdot | \cdot)$ is the likelihood function, $f(\cdot | \cdot)$ is used to denote the joint distributions of the random effects outlined in the previous section, and $\pi(\cdot)$ is used to denote the prior distributions. 

\subsection*{Computational Details}

A Metropolis-within-Gibbs Markov chain Monte Carlo (MCMC) algorithm is used to explore the posterior distribution. Gibbs updates are used for the variance parameters and latent factor covariate effects, and Metropolis-Hastings updates are used for all random effects and the intercepts. Adaptive MCMC \citep{Roberts2009} is used to ensure adequate mixing for all parameters with Metropolis-Hastings updates. The centering constraints of the latent factor $\nu_i$ and the loadings $\alpha_i^D$ are enforced by reparameterizing and centering within the MCMC algorithm. The full conditional distributions for each unknown quantity are provided below, and the code for the MCMC algorithm is included as a supplement. In what follows, $\pi(\cdot | \hdots)$ will be used to denote the full conditional distribution of a random variable given the data and all other random quantities. 

The spatial factor is updated using Metropolis-Hastings updates. To enforce the centering constraint $1/n \sum_i \nu_i=0$, we reparameterize by introducing variables $\nu_i^*$ and updating each $\nu_i^*$ individually then setting $\boldsymbol{\nu}$ equal to the centered $\boldsymbol{\nu}^*$. That is, for $i=1,...,n$, a new value of $\nu_i^*$ is proposed, and the resulting proposed spatial factor is $\boldsymbol{\nu}=\boldsymbol{\nu}^* - \bar{\boldsymbol{\nu}}^*$, where $\bar{\boldsymbol{\nu}}^*$ is the mean of $\boldsymbol{\nu}^*$. The Metropolis-Hastings acceptance ratio is based on the full conditional distribution of $\boldsymbol{\nu}$, given by
\begin{equation*}
\begin{aligned}
\pi(\boldsymbol{\nu}|\hdots) &\propto L(\boldsymbol{\lambda}^T, \boldsymbol{\lambda}^D | \textbf{Y}^T, \textbf{Y}^D) f(\boldsymbol{\nu} | \tau^2),
\end{aligned}
\end{equation*}
where $L(\boldsymbol{\lambda}^T, \boldsymbol{\lambda}^D | \textbf{Y}^T, \textbf{Y}^D)$ is as given in \eqref{eq:lik} and $f(\boldsymbol{\nu} | \tau^2) \propto \exp \left(-1/(2 \tau^2) (\boldsymbol{\nu}-\bf{X} \boldsymbol{\beta})'\bf{Q}(\boldsymbol{\nu}-\bf{X} \boldsymbol{\beta}) \right)$.

As previously described, the spatial loadings for death require a mean one centering constraint. This is accomplished by introducing variables $\alpha_i^*$ that are updated individually, then setting $\boldsymbol{\alpha}^D = \boldsymbol{\alpha}^* - \bar{\boldsymbol{\alpha} }^* +1$. The spatial loadings are updated with Metropolis-Hastings updates based on the full conditional distribution
\begin{equation*}
\pi(\boldsymbol{\alpha}^D|\hdots) \propto L(\boldsymbol{\lambda}^T, \boldsymbol{\lambda}^D | \textbf{Y}^T, \textbf{Y}^D) f(\boldsymbol{\alpha}^D | \tau^2_D),
\end{equation*}
where $f(\boldsymbol{\alpha}^D | \tau^2_D) \propto \exp \left(-(1/2 \tau^2_D) (\boldsymbol{\alpha^D}-\bf{1})'\bf{Q}(\boldsymbol{\alpha^D}-\bf{1})  \right)$.

The independent random errors $\epsilon_i^D, \epsilon_i^T$ are updated for each $i=1,...,n$ according to the full conditional distributions
\begin{equation*}
\pi(\epsilon_i^T, \epsilon_i^D | \hdots) \propto L(\lambda_i^T, \lambda_i^D | Y_i^T, Y_i^D) f(\epsilon_i^T | \sigma^2_T) f(\epsilon_i^D | \sigma^2_D),
\end{equation*}
where $L(\lambda_i^T, \lambda_i^D | Y_i^T, Y_i^D)$ is the $i$th factor in \eqref{eq:lik} and $f(\epsilon_i^T | \sigma^2_T), f(\epsilon_i^D | \sigma^2_D)$ are the zero-mean normal probability density functions with variances $\sigma^2_T, \sigma^2_D$. 

The intercepts $\beta_0^D$, $\beta_0^T$ are updated individually using Metropolis-Hastings updates. The full conditional distributions are given by
\begin{align*}
\pi(\beta_0^D| \hdots) \propto L(\boldsymbol{\lambda}^D | \textbf{Y}^D) \pi(\beta_0^D) \\
\pi(\beta_0^T| \hdots) \propto L(\boldsymbol{\lambda}^T | \textbf{Y}^T) \pi(\beta_0^T),
\end{align*}
where by the conditional independence, $L(\boldsymbol{\lambda}^D | \textbf{Y}^D)$ and $L(\boldsymbol{\lambda}^T | \textbf{Y}^T)$ are simply the factors in \eqref{eq:lik} corresponding to death and treatment, respectively, and by assumption, $\pi(\beta_0^D)=\pi(\beta_0^T)=1$. 

The vector of latent factor regression coefficients $\boldsymbol{\beta}$ can be updated with a Gibbs step. The full conditional distribution is multivariate normal with covariance matrix $\Sigma_{\beta}=\left( 1/\tau^2 {\bf{X}}' {\bf{Q}} {\bf{X}}+1/4 {\bf{I}} \right)^{-1}$ and mean vector $1/\tau^2 \Sigma_{\beta} \bf{X}' \bf{Q} \boldsymbol{\nu}$

The variance parameters $\tau^2, \tau^2_D, \sigma_D^2, \sigma_T^2$ can be updated with Gibbs steps. The prior distributions specified by \eqref{eq:varprior} imply $\pi(\tau^2) \propto \tau^{-1}$, $\pi(\tau_D^2) \propto \tau_D^{-1}$, $\pi(\sigma_D^2)\propto \sigma_D^{-1}$ and $\pi(\sigma_T^2) \propto \sigma_T^{-1}$ \cite{Gelman2006}. Thus, the full conditional distributions are
\begin{align*}
\pi(\tau^2 | \hdots) \propto f(\boldsymbol{\nu} | \boldsymbol{\beta},\tau^2) \pi(\tau^2) \propto  (\tau^2)^{-n/2-1/2} \exp \left( -\frac{1}{2\tau^2} (\boldsymbol{\nu}-\bf{X}\boldsymbol{\beta})'\bf{Q}(\boldsymbol{\nu}-\bf{X}\boldsymbol{\beta}) \right) \\
\pi(\tau^2_D | \hdots) \propto f(\boldsymbol{\alpha}_D | \tau_D^2) \pi(\tau_D^2) \propto  (\tau_D^2)^{-n/2-1/2} \exp \left( -\frac{1}{2\tau_D^2} (\boldsymbol{\alpha}^D-\bf{1})'\bf{Q}(\boldsymbol{\alpha}^D-\bf{1}) \right)\\
\pi(\sigma_D^2 | \hdots) \propto f(\boldsymbol{\epsilon}^D | \sigma_D^2) \pi(\sigma_D^2) \propto (\sigma_D^2)^{-n/2-1/2} \exp \left( -\frac{1}{2\sigma_D^2} \boldsymbol{\epsilon_D}'\boldsymbol{\epsilon_D} \right) \\
\pi(\sigma_T^2 | \hdots) \propto f(\boldsymbol{\epsilon}^T | \sigma_T^2) \pi(\sigma_T^2) \propto (\sigma_T^2)^{-n/2-1/2} \exp \left( -\frac{1}{2\sigma_T^2} \boldsymbol{\epsilon_T}'\boldsymbol{\epsilon_T} \right),
\end{align*}
which we can identify as being from the inverse-gamma family of distributions with shape parameters $n/2-1/2$ and scale parameters of $1/2 (\boldsymbol{\nu}-\bf{X}\boldsymbol{\beta})'\bf{Q}(\boldsymbol{\nu}-\bf{X}\boldsymbol{\beta})$, $1/2 (\boldsymbol{\alpha}^D-\bf{1})'\bf{Q}(\boldsymbol{\alpha}^D-\bf{1})$, $1/2 \boldsymbol{\epsilon}_D'\boldsymbol{\epsilon}_D$, and $1/2 \boldsymbol{\epsilon}_T'\boldsymbol{\epsilon}_T$, respectively. 

\newpage 

\subsection*{Supplemental Results}
\setcounter{figure}{0}   
\setcounter{table}{0}

\begin{table}
	\caption{\label{table:trt_codes} Diagnostic codes used to define treatment}
	\fbox{%
		\begin{tabular}{ll}
			\multicolumn{2}{l}{\textbf{ICD-9 Diagnostic Codes}}\\
			304.00 & Opiate Type Dependence, Unspecified Use\\
			304.01 & Opioid Type Dependence, Continuous Use\\
			304.02 & Opioid Type Dependence, Episodic Use\\
			304.03 & Opioid Type Dependence, in Remission\\
			304.70 & Combinations of Opioid Type Drug with Any Other, Unspecified Use\\
			304.71 & Combinations of Opioid Type Drug with Any Other, Continuous Use\\
			304.72 & Combinations of Opioid Type Drug with Any Other, Episodic Use\\
			304.73 & Combinations of Opioid Type Drug with Any Other, in Remission\\
			305.50 & Opioid Abuse, Unspecified Use\\
			305.51 & Opioid Abuse, Continuous Use\\
			305.52 & Opioid Abuse, Episodic Use\\
			305.53 & Opioid Abuse, in Remission\\
			& \\
			\multicolumn{2}{l}{\textbf{DSM-IV-TR Diagnostic Codes}}\\
			292.00 & Opioid Withdrawal \\
			292.11 & Opoid-Induced Psychotic Disorder, with Delusions\\
			292.12 & Opoid-Induced Psychotic Disorder, with Hallucinations\\
			292.81 & Opioid Intoxication Delirium\\
			292.84 & Opioid-Induced Mood Disorder\\
			292.89 & Opioid Intoxication\\
			292.89 & Opioid-Induced Sexual Dysfunction\\
			292.89 & Opioid-Induced Sleep Disorder\\
			292.90 & Opioid-Related Disorder, NOS\\
			304.00 & Opioid Dependence \\
			305.50 & Opioid Abuse\\
		\end{tabular}}
	\end{table} 
	
	\begin{figure}[t]
		\centering
		\subfigure[Observed Death Rates]{
			\label{fig:obs_death}
			\includegraphics{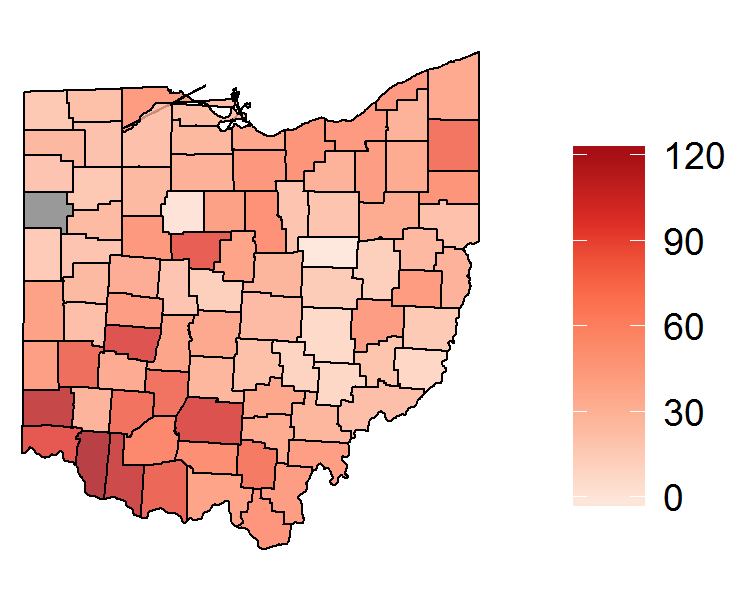}
		}
		\subfigure[Observed Treatment Rates]{
			\label{fig:obs_trt}
			\includegraphics{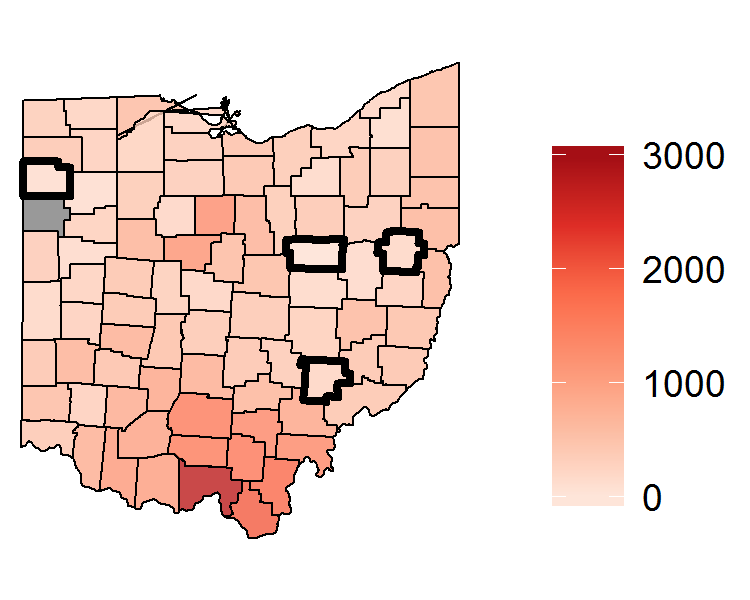}
		}
		\subfigure[Scatterplot of Observed Treatment and Death Rates]{
			\label{fig:obs_scatter}
			\includegraphics{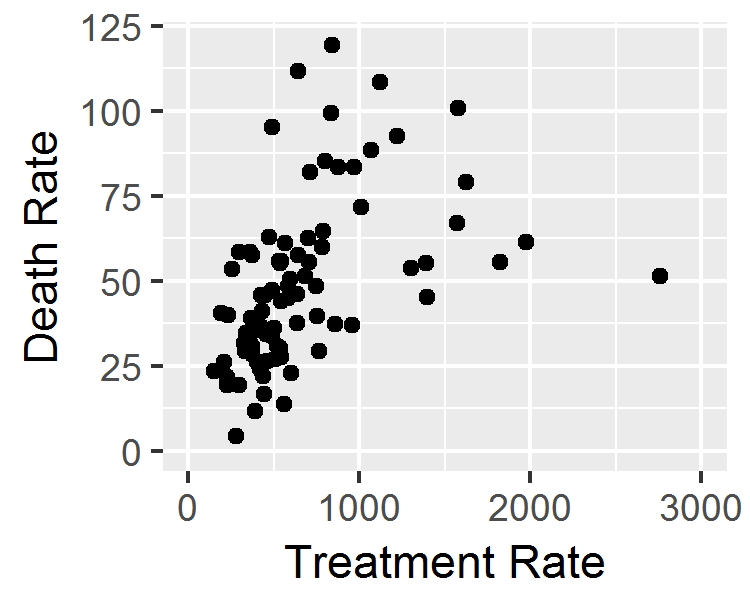}
		}
		\caption[Observed Data]{Observed County Opioid Associated Death Rates and Treatment Admission Rates per 100,000 Residents, Ohio Counties, 2013-2015}
		\label{fig:obs}
	\end{figure}
	
	\begin{figure}[b]
		\centering
		\includegraphics{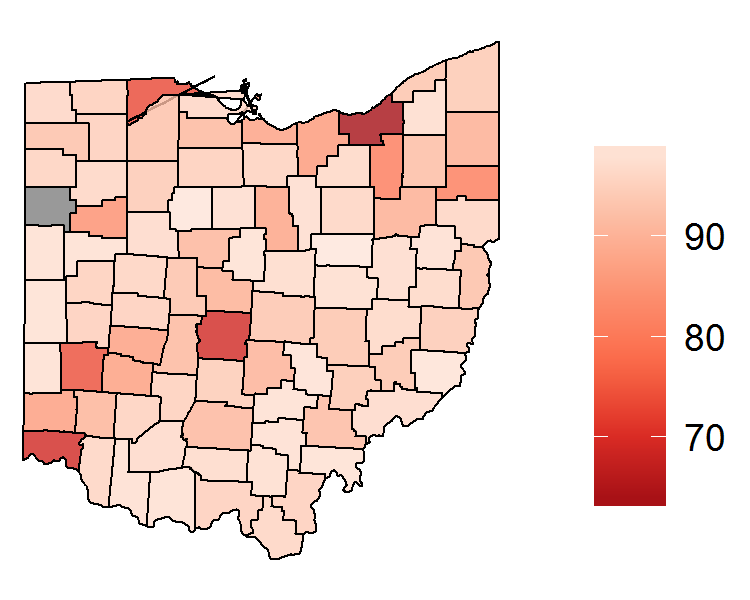}
		\caption[Proportion of White Residents]{Estimated Proportion of White Residents for Each County}
	\end{figure}
	
	\begin{figure}
		\centering
		\subfigure[Independent Latent Factor for Death]{
			\label{fig:est_eps_death}
			\includegraphics{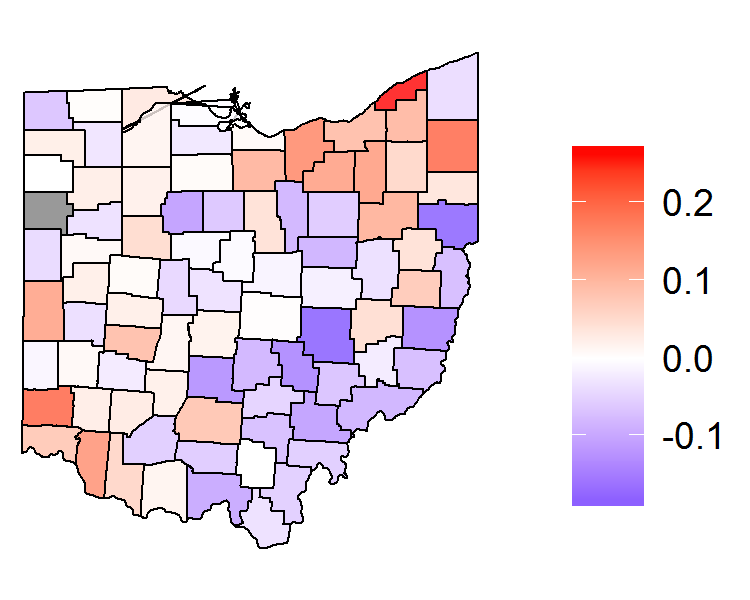}
		}
		\subfigure[Independent Latent Factor for Treatment]{
			\label{fig:est_eps_trt}
			\includegraphics{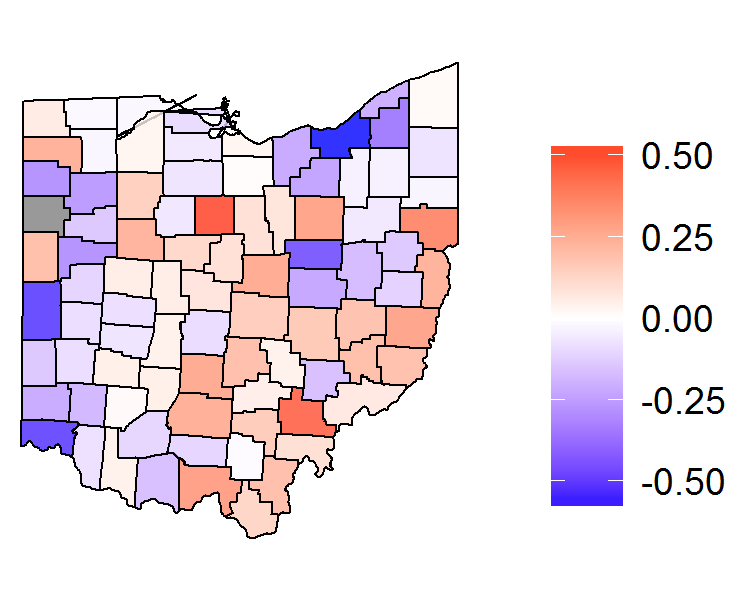}
		}
		\caption[Estimated Independent Latent Factor]{Estimated Independent Latent Factor for Each County for Death and Treatment Rates}
		\label{fig:est_eps}
	\end{figure}

	\begin{table}[b]
		\caption{\label{table:var_effects}Posterior Mean Estimates and 95\% Credible Intervals for Variance Parameters in the Model}
		\centering
		\begin{tabular}{lcc}
			\hline
			Parameter & Posterior Mean & 95\% Credible Interval\\
			\hline
			$\tau^2$&$0.3792$&$(0.1422, 0.7474)$\\
			$\tau^2_D$&$0.2679$&$(0.1488, 0.4341)$\\
			$\sigma^2_D$&$0.0346$&$(0.0012, 0.0846)$\\
			$\sigma^2_T$&$0.00860$&$(0.0438, 0.1428)$\\
			\hline
		\end{tabular}
	\end{table}

\end{document}